\documentclass[aps,prd,showpacs,preprint,showpacs,showkeys,preprintnumbers,superscriptaddress,amsmath,amssymb,nofootinbib]{revtex4-1}

\usepackage{color}
\usepackage{latexsym}
\usepackage{amsmath}
\usepackage{amssymb}
\usepackage[dvips]{graphicx}
\usepackage{color}
\usepackage{eufrak}
\usepackage{euscript}
\usepackage{pstricks}
\usepackage{graphics}
\usepackage{graphicx}
\usepackage{picture}
\def\[{\left\lbrack}
\def\]{\right\rbrack}

\def\({\left(}
\def\){\right)}

\newcommand{\bee}{\begin{equation}}
\newcommand{\eee}{\end{equation}}
\newcommand{\eaa}{\end{eqnarray}}
\newcommand{\baa}{\begin{eqnarray}}

\def\ni{\noindent}

\begin{document}

\title{\Large{Faddeev-Jackiw approach of the noncommutative spacetime Podolsky electromagnetic theory}}

\author{Everton M. C. Abreu}\email{evertonabreu@ufrrj.br}
\affiliation{Grupo de F\' isica Te\'orica e Matem\'atica F\' isica, Departamento de F\'{i}sica, Universidade Federal Rural do Rio de Janeiro, 23890-971, Serop\'edica - RJ, Brazil}
\affiliation{Departamento de F\'{i}sica, Universidade Federal de Juiz de Fora, 36036-330, Juiz de Fora - MG, Brazil}
\author{Rafael L. Fernandes}\email{rlfernandes@fisica.ufjf.br}
\affiliation{Departamento de F\'{i}sica, Universidade Federal de Juiz de Fora, 36036-330, Juiz de Fora - MG, Brazil}
\author{Albert C. R. Mendes}\email{albert@fisica.ufjf.br}
\affiliation{Departamento de F\'{i}sica, Universidade Federal de Juiz de Fora, 36036-330, Juiz de Fora - MG, Brazil}
\author{Jorge Ananias Neto}\email{jorge@fisica.ufjf.br}
\affiliation{Departamento de F\'{i}sica, Universidade Federal de Juiz de Fora, 36036-330, Juiz de Fora - MG, Brazil}

\date{\today}

\begin{abstract}
\ni The interest in higher derivatives field theories has its origin mainly in their influence concerning the renormalization properties of physical models and to remove ultraviolet divergences.
The noncommutative Podolsky theory is a constrained system that cannot by directly quantized by the canonical way. In this work we have used the Faddeev-Jackiw method in order to obtain the Dirac brackets of the NC Podolsky theory.
\end{abstract}



\maketitle


\section{Introduction}

The first known paper describing a noncommutative (NC) space-time was published by Snyder in 1947 \cite{snyder}.
His work was motivated by the necessity to tame the divergences that haunt QFT such as quantum electrodynamics, to mention an example. 
The idea of constructing a NC space-time had its inspiration in quantum mechanics.   It was introduced for the first time by Heisenberg where, in a quantum phase space, the coordinate operators  $\hat{x}^i$ satisfy a kind of uncertainty relations
$[\hat{x}^i,\hat{x}_j]=i\,\theta^i_j$.
The objective was to use a  space-time coordinate system with a NC structure in small scales and, in this way, to introduce a cutoff in the ultraviolet regime. However, shortly after Snyder's paper, C. N. Yang \cite{yang} demonstrated that, alas, even using this NC algebra, the QFT divergences are still there.

This result has put Snyder's noncommutativity (NCY) in complete sleep for more than 50 years until Seiberg and Witten (SW) \cite{sw} demonstrated that the algebra resulting from a string model embedded in a magnetic field has demonstrated itself to have a NC algebra. Since then we have seen an prolific research concerning several NC formulations that deserves our attention and profound analysis. In other words, SW have shown that QFT described in a NC space-time can be obtained as a limit of string theory. Their effects can be systematically analyzed as a perturbation using the so-called SW map. A gauge theory described in a NC space-time can be treated as usual gauge theories with the same degrees of freedom.   But, with an additional  $\theta$-parameter term. In other words, the SW map allows us to construct a minimal action principle of usual commutative fields, so the effective Lagrangian is expanded as series in original terms plus additional  $\theta$-parameter terms, as we have just said.

Currently, we have some theoretical reasons that could justify to introduce NCY concerning the position coordinates. As an example, one of the great challenges in theoretical physics is to unify, in a single and consistent framework,   quantum mechanics and general relativity. The combination of special relativity and QFT was already accomplished through the Klein-Gordon and Dirac models. The introduction of a Planck scale quantity such as the $\theta$-parameter into the space-time would be one way to introduce a quantum ingredient into the classical space-time.   However, the way to combine  general relativity with the quantum mechanics formalism is still mysterious, but NCY is considered a way to accomplish the task.  

The analysis of constrained systems has attracted a great interest in theoretical physics since the works of Dirac and Bergman  \cite{dirac}. Several years later,
Faddeev and Jackiw (FJ)  \cite{fj} obtained a general method to study constrained systems without the Dirac's classification in first or second class constraints.   The FJ technique, also called as symplectic method, can be used as a geometric way to obtain the Dirac brackets.  The advantage of this method, as we said, is that it is not necessary to classify the constraints into the so-called primary or secondary, first-class or second-class and some ambiguities mentioned above could be avoided.   After the work of FJ, J. Barcelos Neto and C. Wotzasek \cite{bw}, have showed that a symplectic method can be extended in such a way that the constraints can be incorporated.  

The FJ's proposal to treat constrained systems in order  to eliminate the superfluous degrees of freedom,  cannot always be done.  The purpose of this paper is to discuss this problem in an alternative way, namely we will apply the FJ
method \cite{fj} to the NC space-time Podolsky electromagnetic theory. 

Podolsky electrodynamics has been introduced with the objective of eliminating the divergence that appears in Maxwell electrodynamics at short distances. Consequently, in Podolsky electrodynamics the energy associated with a point particle is finite \cite{podolsky}. The treatment of the Podolsky electrodynamic theory on NC space-time was developed in \cite{afma} where it was analyzed the gauge invariance of this model. The Moyal product was introduced in Podolsky's Lagrangian as well as the SW map. The NC version of Podolsky electrodynamics is a constrained system. The FJ analysis of this higher-derivative model was accomplished in \cite{amnowx}.  In this paper we have analyzed both scenarios, namely, we have used the FJ method to analyze a NC version of the Podolsky higher-derivative electrodynamics.


This paper follows such organization in a way that in section II we have described the main steps of the Faddeev-Jackiw method.  In section III we have discussed some results of  the NC space-time Podolsky electrodynamics.  In section IV we have applied the FJ method to the NC version of Podolsky's electrodynamics to obtain the Dirac brackets, which means quantization.  We have left the final discussions and the conclusions for the final section.


\section{The Faddeev-Jackiw method}

Let us begin with a brief description of the symplectic formulation where we can define a compact notation concerning the position coordinates and momenta of our phase space. In this way, we will define a quantity $Y^\alpha$, where $\alpha=1,2,\ldots,2N$, as a set of coordinate and momenta of phase space with $2N$ bosonic degrees of freedom, such as
\begin{align}
Y^i&=q_i,\nonumber\\
Y^{N+i}&=p_i \qquad (i=1,2,\ldots,N)\,\,,
\end{align}

\ni where $q_i$ are the position coordinates and $p_i$, the momenta.
With these last relations, it is easy to see that the Poisson brackets take the compact form
\begin{equation}
\{Y^\alpha,Y^\beta\}=\epsilon^{\alpha\beta},
\label{eq_16}
\end{equation}

\ni where $\epsilon^{\alpha\beta}$ is a matrix given by
\begin{equation}
(\epsilon^{\alpha\beta})=\left(\begin{array}{cc}
\mathbb{O}&\openone\\
-\openone&\mathbb{O}
\end{array}\right),
\end{equation}

\ni where $\mathbb{O}$ is a zero matrix and $\openone$ is the identity matrix.   Both have $N{\times}N$ dimension.

In the analysis of a constrained system, the generalization of Eq. $(\ref{eq_16})$ is given by
\begin{equation}
\{Y^\alpha,Y^\beta\}=f^{\alpha\beta},
\label{eq_16.5}
\end{equation}

\ni where $f^{\alpha\beta}$ is a antisymmetric tensor.

Let us start with a first-order time derivative Lagrangian,   
\begin{equation}
L=a_\alpha(Y)\dot{Y}^\alpha-V(Y),
\label{eq_17}
\end{equation}
where $V(Y)$ is the potential.

Using the Euler-Lagrange equation
\begin{equation}
\frac{d}{dt}\left(\frac{\partial{L}}{\partial\dot{Y}^\alpha}\right)-\frac{\partial{L}}{\partial{Y^\alpha}}\,=\,0,
\end{equation}

\ni and from the Lagrangian in \eqref{eq_17}, we have that
\begin{equation}
f_{\alpha\beta}\dot{Y}^\beta=\partial_\alpha{V}\,\,,
\label{eq_18}
\end{equation}

\ni where $\partial_\alpha=\partial /\partial{x}_\alpha$ and the symplectic tensor is given by
\begin{equation}
f_{\alpha\beta}=\frac{{\partial}a_\beta}{{\partial}Y_\alpha}-\frac{{\partial}a_\alpha}{{\partial}Y_\beta}\,\,.
\label{eq_18.5}
\end{equation}

If $f_{\alpha\beta}$ is not a singular matrix, we can rewrite Eq. $(\ref{eq_18})$ using  its inverse $f^{\alpha\beta}$ such that
\begin{equation}
\dot{Y}^\alpha=f^{\alpha\beta}\partial_\beta{V}\,\,,
\label{eq_19}
\end{equation}

\ni which are the equations of motion. However, if  $\det (f_{\alpha\beta})\neq0$ we have that $f_{\alpha\beta}$ is singular and we cannot write the velocities as in $(\ref{eq_19})$.

In order to describe the symplectic formalism with constraints \cite{bw}, let us denote $f_{\alpha\beta}$ as $f^{(0)}_{\alpha\beta}$ in such a way that it contains $M$ constraints, where  $M<2N$, and $v^{(0)}_m$ are the zero modes and $m=1,2,\ldots,M$. So,
\begin{equation}
f^{(0)}_{mn}\,v^{(0)}_{mn}=0\,\,,
\end{equation}

\ni and, from Eq. $(\ref{eq_18})$, we can write that
\begin{equation}
v^{(0)}_{m}\,\partial_m{V}=0\,\,,
\end{equation}

\ni which is a constraint.

In order to obtain a twisted form of the tensor $f^{(0)}$ \cite{bw}  we will introduce the constraints into the kinetic part of the Lagrangian in Eq. $(\ref{eq_17})$.   This introduction of these constraints will be carried out by considering the time derivative constraint. Then, we will include  these results into the action using a Lagrangian multiplier.   Namely, the modification obtained is an enlargement of the configuration space and hence, the true symplectic variables are  $(Y^\alpha,\lambda^{(0)}_m)$ and the new Lagrangian will be given by
\begin{equation}
L^{(0)}=a^{(0)}_\alpha(Y)\,\dot{Y}^\alpha+\lambda^{(0)}_m\,\dot{\Omega}^{(0)}_m-V^{(0)}(Y)\,\,,
\end{equation}

\ni or
\begin{equation}
L^{(0)}=\,\Big(a^{(0)}_\alpha(Y)+\lambda^{(0)}_m\,\partial_\alpha\Omega^{(0)}_m \Big)\,\dot{Y}^\alpha-V^{(0)}(Y)\,\,.
\label{eq_19.5}
\end{equation}

From the Lagrangian $(\ref{eq_19.5})$, we can identify the vectors
\begin{align}
a^{(1)}_\alpha&=a^{(0)}_{\alpha}+\lambda^{(0)}_m\partial_\alpha\Omega^{(0)}_m,\nonumber\\
a^{(1)}_m&=0,
\end{align}

\ni and from Eq. $(\ref{eq_18.5})$ we can compute the symplectic tensor
\begin{align}
f^{(1)}_{\alpha{m}}=\,-\partial_ma^{(1)}_\alpha\,\,\qquad
\mbox{and} \qquad
f^{(1)}_{mn}=\,0 \,\,.
\end{align}

This last procedure must be repeated as many times as necessary, until $\det (f)\neq{0}$. When this occurs, the result is the symplectic tensor and consequently the Dirac bracket.


\section{Podolsky's electrodynamics on non-commutative space-time}

\subsection{Noncommutativity in space-time: a few words}

There are some different approaches of NCY concerning the feature of the $\theta$-parameter.  We can have NC algebra,
\bee
\label{B}
\Big[\hat{x}^\mu,\hat{x}^\nu \Big] = i\,\theta^{\mu\nu} \,\,, \qquad \Big[\hat{x}^\mu,\hat{p}_\nu \Big] = i\,\delta^\mu_\nu \,\,, \qquad \Big[\hat{p}_\mu,\hat{p}_\nu \Big] = 0\,\,,
\eee

\ni  where  these commutators indicate operators in Snyder's formulation and the $\theta$-parameter is constant, which is known as the canonical NCY.  There are other approaches where the $\theta$-parameter is not constant \cite{dfr,an}, something like $\theta=\theta(x)$ or $\theta$ is a coordinate of the space-time \cite{dfr,alexei}.   Notice that we have talked only about NC formalism concerning the $\theta$-parameter.   There are many other approaches that consider vriations of the results of the commutators in Eq. \eqref{B}.

One of the most popular approaches of NC theories \cite{reviews} is the one ruled by the well known Moyal-Weyl product \cite{rsh}.  In this approach, the standard product of two NC objects is substituted by the so-called star-product given by

\begin{equation}
\label{A}
\widehat{f}(x) * \widehat{g}(x)= \exp\Big(\frac i2 \theta^{\mu\nu} \partial_{\mu}^x \partial_{\nu}^y \Big)\,\widehat{f}(x) \widehat{g}(y) \Big|_{x=y},
\end{equation}

\ni where  $\theta^{\mu\nu}$ is the NC parameter, of course.


However, the Moyal-Weyl product must have the associative property which cannot be lost, the consequence in this case is that it would not be considered a star-product, which is another issue in the construction of non-constant $\theta$-parameter.  But in our case analyzed here we will use that $\theta^{\mu\nu}$ is constant.  In this case the Moyal-Weyl product is a star-product with the associative property. The constant feature of $\theta^{\mu\nu}$ brings another problem, since we have a fixed direction defined in the right hand side of Eq. (\ref{B}).  This fact breaks the Lorentz invariance of the theory.  The solution of this problem leads us to another approach of the NC theory formulated by Doplicher, Fredenhagen and Roberts (DFR) \cite{dfr}, which is based in general relativity and quantum mechanics arguments.  In our formulation we will analyze the NC version of Podolsky model with a chosen direction.


\subsection{The NC space-time Podolsky model}
 
Let us begin with the action of NC Podolsky field which is given by
\begin{equation}
\label{1}
S=\int \Big(-\frac 14 \widehat{F}_{\mu\nu} * \widehat{F}^{\mu\nu}+\frac 12\,a^2\partial_{\lambda}\widehat{F}^{\lambda\nu} * \partial^{\mu}\widehat{F}_{\mu\nu} \Big)d^4 x\,\,,
\end{equation}

\ni where, as explained above, $*$ means the Moyal-Weyl product.  The hat notation indicates NC space-time space-time objects and $\widehat{A}_{\mu}$ and $\widehat{F}_{\mu\nu}$ are the NC vector potential and field strength tensor respectively.  The term $a$ represents $1/m$ which is the inverse mass of the $\widehat{A}_{\mu}$ field.  We will use $(-+++)$  and the summation notation for repeated indices is assumed from now on.

Since we know that for two objects within an integral \cite{rsh} the Moyal-Weyl product can be substituted by an ordinary product, we can write Eq. (\ref{1}) as
\begin{equation}
\label{2}
S=\int \Big(-\frac 14 \widehat{F}_{\mu\nu}  \widehat{F}^{\mu\nu}+\frac 12{a^2}\partial_{\lambda}\widehat{F}^{\lambda\nu}  \partial^{\mu}\widehat{F}_{\mu\nu} \Big)d^4 x\,\,,
\end{equation}

\ni and the action just above describes the Podolsky theory in a NC space-time.  In order to show the NC 
terms, i.e., $\theta$-terms, of this model we have to use the well known SW map \cite{sw}, which tells us that
\begin{equation}
\label{3}
\widehat{A}_{\mu}=A_{\mu}-\frac 12 \,\theta^{\alpha\beta} A_{\alpha}(\partial_{\beta} A_{\mu}+F_{\beta\mu})
\end{equation}

\ni and consequently
\begin{equation}
\label{4}
\widehat{F}_{\mu\nu}=F_{\mu\nu}+\theta^{\rho\sigma}(F_{\mu\rho} F_{\nu\sigma}-A_{\rho} \partial_{\sigma} F_{\mu\nu} )\,\,,
\end{equation}

\ni where $\theta^{\mu\nu}$ is the  constant NC parameter explained above and the Moyal-Weyl product has the star-products associative property. 

Under the SW map, at the lowest non-trivial order, the duality and/or the equivalence relations are kept unchanged. Besides, the SW map does not distinguish between the unequal gauge invariances.
Substituting Eqs. (\ref{3}) and (\ref{4}) into Eq. (\ref{2}) we have the NC Podolsky Lagrangian density with the same degrees of freedom but with additional terms due to the NC parameter first-order 

\begin{align}
\label{5}
\widehat{\cal L}_{Pod}&=-\frac 14F^2-\frac 18\,\theta^{\rho\sigma}\Big(4 F^{\mu}_{\:\:\:\rho} F^{\nu}_{\:\:\:\sigma} F_{\mu\nu}-\, F^2 F_{\rho\sigma} \Big) \nonumber \\
&+\frac 12{a^2}\partial_{\lambda}F^{\lambda\nu} \Big[\partial^{\mu} F_{\mu\nu}+2 \theta^{\rho\sigma} \partial^{\mu} \Big(F_{\mu\rho} F_{\nu\sigma}- A_{\rho} \partial_{\sigma} F_{\mu\nu} \Big) \Big]\,\,,
\end{align}

\ni where we can see quickly that when $\theta^{\mu\nu}=0$ we can recover the standard commutative model. From now on we will assume that making such $\theta$ approximation ($\theta=0$) in any resulting NC equation, we certainly recover the commutative results. 

We can write the above NC  Lagrangian in Eq. $\eqref{5}$, in terms of the fundamental electromagnetic fields such as
\begin{align}
\widehat{\mathcal{L}}_{Pod}&=\frac{1}{2}(E^2-B^2)+\frac{1}{2}a^2\left[(\nabla\cdot\vec{E})^2-(\dot{\vec{E}}-\nabla\times\vec{B})^2\right]-\left[(\vec{\theta}\cdot\vec{E})(\vec{B}\cdot\vec{E})-(\vec{\theta}\cdot\vec{B})E^2\right]\nonumber\\
&+\frac{1}{2}(B^2-E^2)\vec{\theta}\cdot\nabla\times\vec{A}+a^2\Big\{\left[(\vec{\theta}\cdot\dot{\vec{B}})\vec{E}+(\vec{\theta}\cdot\vec{B})\dot{\vec{E}}\right]\cdot\nabla\times\vec{B}-\vec{\theta}\cdot(\nabla\times\vec{B})\partial_0(\vec{E}\cdot\vec{B})\nonumber\\
&+\left[\vec{\theta}\cdot(\vec{A}\times\nabla)\right]\vec{E}\Big\}\,\,.
\label{5.1}
\end{align}

\ni where we have used the relation $\theta^{ij}=\epsilon^{ijk}\theta_k$ and $\vec{\theta}=(\theta_1,\theta_2,\theta_3 )$.

Considering the equations of motion, we have the generalized form for a higher-derivative model,
\begin{equation}
\label{5.1a}
\frac{\partial\widehat{\cal L}_{Pod}}{\partial A_\lambda}-\partial_\mu \frac{\partial \widehat{\cal L}_{Pod}}{\partial (\partial_\mu A_\lambda )} + \partial_\mu \partial_\nu \frac{\partial \widehat{\cal L}_{Pod}}{\partial (\partial_\mu \partial_\nu A_\lambda )} - \partial_\alpha \partial_\mu \partial_\nu \frac{\partial \widehat{\cal L}_{Pod}}{\partial ( \partial_\alpha \partial_\mu \partial_\nu A_\lambda )}\,=\,0
\end{equation} 

\ni and hence, the equations of motion for the NC Podolsky model are
\begin{align}
&(1+a^2\Box)\partial_{\rho}F^{\rho\lambda}+\partial_\rho\Big\{\frac{1}{2}\left[\theta^{\mu{[\rho}}F^{\lambda]\nu}F_{\nu\mu}+\theta^{\mu[\rho}F^{\lambda]\nu}F_{\mu\nu}+\theta^{\beta\alpha}F^{\,\,\,[\rho}_{\beta}F^{\lambda]}_\alpha\right]\nonumber\\
&-\frac{1}{4}\Big[\theta^{\beta\alpha}F^{\rho\lambda}F_{\beta\alpha}+\theta^{\rho\lambda}F^2\Big]+a^2\left[\theta^{\beta[\rho}\partial^{\alpha]}
F^{\lambda}_{\:\:\:\alpha}\partial^{\mu}F_{\mu\beta}+\theta^{\alpha[\rho}\partial^{{\lambda}]}F_{\nu\alpha}\partial_{\beta}F^{\beta\nu}
+\theta^{\rho\lambda}\partial_\alpha{F}^{\alpha}_{\:\:\:\nu}\partial_{\beta}F^{\beta\nu}\right]\Big\}\nonumber\\
+&\partial_\rho\partial_\sigma\Big[a^2F^\sigma_{\:\:\:\beta}\theta^{\beta[\rho}\partial_\alpha{F}^{\lambda]\alpha}+\theta^{\beta\sigma}\partial_\alpha{F}^{\alpha[\lambda}\partial^{\rho]}A_\beta\Big]=0\,\,.
\label{5.2}
\end{align}

Thus, analyzing Eq. \eqref{5} we can calculate the momenta relative to the phase space described by the following pairs, $(A_\mu\, , \rho^\mu ),\,(\dot{A}_\mu \,, \pi^\mu)$ and $(\ddot{A}_\mu \, , \chi^\mu)$, provided by the coordinates and their respective momenta in the phase space.  The momenta for higher-derivative models are given by
\begin{align}
\rho^\mu=&\frac{\partial\mathcal{L}}{\partial\dot{A}_\mu}-\partial_0\frac{\partial\mathcal{L}}{\partial\ddot{A}_\mu}-2\,\partial_n\frac{\partial\mathcal{L}}{\partial(\partial_n\dot{A}_\mu)}+3\,\partial_0\partial_n\frac{\partial\mathcal{L}}{\partial(\partial_n\ddot{A}_\mu)}+3\,\partial_n\partial_m\frac{\partial\mathcal{L}}{\partial(\partial_n\partial_m\dot{A}_\mu)}+\partial_0\partial_0\frac{\partial\mathcal{L}}{\partial(\dddot{A}_\mu)},\nonumber\\
\pi^\mu=&\frac{\partial\mathcal{L}}{\partial\ddot{A}_\mu}-3\,\partial_n\frac{\partial\mathcal{L}}{\partial\partial_n\ddot{A}_\mu}-\partial_0\frac{\partial\mathcal{L}}{\partial\ddot{A}_\mu},\nonumber\\
\chi^\mu=&\frac{\partial\mathcal{L}}{\partial\dddot{A}_\mu}\,\,.
\label{5.3}
\end{align}

It is easy to see that, since ${\mathcal L}_P$ does not have a $\dddot{A}_\mu$, all the relative derivatives are zero.   Hence, we have explicitly that 
\begin{align}
\chi^{\zeta}&=0,\nonumber\\
\pi^0&=0,\nonumber\\
\pi^n&=a^2\partial_{\alpha}F^{\alpha{n}}+a^2\theta^{ij}\partial^{\alpha}(F_{\alpha{i}}F^n_j-A_i\partial_jF_{\alpha}^n)+a^2\theta^{nj}(\partial_{\alpha}F^{\alpha{k}})F_{kj}+a^2\theta^{ij}\partial_j(A_i\partial_{\alpha}F^{\alpha{n}}),\nonumber\\
\rho^0&=a^2\partial_n(\partial_{\alpha}F^{\alpha{n}})+a^2\theta^{ij}\partial_n\partial^{\alpha}(F_{\alpha{i}}F^{n}_j-A_i\partial_jF_{\alpha}^n)+a^2\theta^{nj}\partial_n(\partial_{\alpha}F^{\alpha{k}}F_{kj})+a^2\theta^{ij}\partial_{n}\Big[\partial_j(A_i\partial_{\alpha}F^{\alpha{n}})\Big],\nonumber\\
\rho^n&=-F^{0n}(1-\frac{1}{2}\theta^{ij}F_{ij})+2a^2\partial^n\partial_iF^{i0}-a^2\partial_0(\partial_{\alpha}F^{{\alpha}n})-\theta^{nj}F_{0k}F^k_j-\theta^{ij}F_{0i}F^n_j\nonumber\\
&-a^2\Big[\theta^{nj}\big(\partial_{\alpha}F^{{\alpha}m}\partial_jF_{0m}-\partial_k(\partial_0F^{0k}F_{0i}-\partial^l(\partial_kF^{k0}F_{li})+\partial_m(\partial_kF^{km}F_{0i}-\partial_0(\partial_{\alpha}F^{{\alpha}k}F_{kj}\big)\Big]\nonumber\\
&+\theta^{ij}\Big\{-\partial_j(\partial_0F^{on}F_{0i})-\partial_j(\partial_kF^{0n}\partial_0A_i)+\partial_j(\partial_kF^{k0}\partial^nA_i)+\partial_j(\partial_kF^{kn}F_{0i}-\partial_j(\partial_kF^{kn}\dot{A}_i)\nonumber\\
&-\partial^n(F_{li}\partial^lF_{0j}-\partial^lA_i\partial_jF_{l0})-\partial_0\Big[\partial^{\alpha}(F_{{\alpha}i}F^n_j-A_i\partial_jF_\alpha^n)\Big]-3\partial_0\partial_j(A_i\partial_0F^{0n})
+3\partial^n\partial_j\Big[A_i\partial_kF^{k0})\Big] \Big\}\,\,.
\label{5.4}
\end{align}

\ni and the canonical Hamiltonian density is given  by
\begin{equation}
\label{5.5}
\widehat{\cal H}_{Pod}=\pi ^n \ddot{A}_n+\rho^0 \dot{A}_0+\rho^n \dot{A}_n- \widehat{\cal L}_{Pod}\,\,,
\end{equation}

\ni where the momenta are given above and the Lagrangian is written in Eq. $\eqref{5.1}$. Notice, that in our case, a higher derivative theory, the phase space is given by $(A_0\,, A_n \,, \dot{A}_0\, , \dot{A}_n \,, \rho_0\, , \rho_n\, , \pi_0 \,,\pi_n)$.
Hence, substituting these values in the equation above we have that
\begin{align}
\widehat{\mathcal{H}}_{Pod}&=\rho^{\mu}\dot{A}_{\mu}-\frac{({\pi^n})^2}{2a^2}(1-\theta^{ij}\partial_jA_i)-\pi_n\partial^n\dot{A}^0+\pi_n\partial_iF^{in}\nonumber\\
&+\frac{1}{4}(2F^{0k}F_{0k}+F^{kl}F_{kl})(1-\frac{1}{2}\theta^{ij}F_{ij})+
\frac{\theta^{ij}}{2}\Big(F_{0i}F_{0k}F^{k}_{\:\:\:j}+F^{k}_{\:\:\:i}F_{k0}F_{0j}+F_{kl}F^k_{\:\:\:i}F^{l}_{\:\:\:j}\Big)\nonumber\\
&-\frac{a^2}{2}(\partial^{k}\dot{A}_k-\partial_k^2 A_0)^2
-a^2\theta^{ij}\Big[\partial_{\lambda}F^{\lambda{0}}\partial^{n}(F_{n{i}}F_{0{j}}-A_i\partial_jF_{n{0}})\Big] \nonumber \\
&+\theta^{ij}\Big[\pi^n\partial^{\mu}(F_{\mu{i}}F_{n{j}}-A_i\partial_jF_{\mu{n}})\Big],
\end{align}

\ni where, in order to keep it simple, we have chosen to keep the stress tensor $F_{ij}$.  In the next section we will discuss the quantization of the Lagrangian in Eq. \eqref{5.1} through the FJ formalism.  

\section{Faddeev-Jackiw quantization and the noncommutative Podolsky electrodynamics}

FJ have demonstrated that the description of constrained systems by its canonical, first-order form, can provide a considerable improvement into the Dirac-Bergmann's approach for this area of research.   Considering symplectic manifolds, the geometrical structure, well known as Dirac (or generalized) bracket, can be obtained directly from the inverse of the nonsingular symplectic two-form matrix.  For nonsymplectic manifolds, this two-form is degenerated and its inverse cannot be obtained to provide the generalized brackets.   This singular feature of the symplectic matrix characterizes the existence of constraints that have to be considered with care in order to yield consistent results.   

At this point we have two ways to deal with this problem, one of them is through Dirac's approach, which determines that we have to introduce the constraints into the potential part, the Hamiltonian, of the canonical Lagrangian, and consequently we would have the final form of Dirac brackets.   These last ones are consistent with the constraints and can be mapped into quantum commutators.   

The second way we have is the point of view of FJ, which includes the constraints directly into the canonical part of the first-order Lagrangian which leads to an invertible two-form symplectic matrix from where the Dirac brackets are directly obtained.   
In this section we will apply this symplectic formalism to find the Dirac brackets for the NC Podolsky theory. The Dirac brackets will be found by this geometric way, as was discussed before.

The Lagrangian in Eq. (\ref{5.1}) obviously have higher order terms, so, we have to make use of the Ostrogradski approximation \cite{Ostrogradski,Weiss,Chang} to work with these terms. Hence, we have to introduce two new sets of canonical pairs, $(\Gamma^\mu\equiv\dot{A}^\mu,\pi^\mu)$ and $(\phi^\mu\equiv\ddot{A}^\mu,\chi^\mu)$. In this way, we can write
\begin{align}
(\rho^{\mu},{A}^{\mu})\,\,, \qquad
(\pi^{\mu},\Gamma^{\mu})\,\, \qquad \mbox{and} \qquad
(\phi^{\mu},\chi^\mu)\,\,.
\label{5.6}
\end{align}

Using Eq. $(\ref{5.6})$ we can rewrite the Lagrangian $(\ref{5.1})$ in a first order form
\begin{equation}
\widehat{\mathcal{L}}=\pi_\mu\dot{\Gamma}^\mu+\rho_\mu\dot{A}^\mu-\widehat{V}^{(0)},
\label{eq_42}
\end{equation}

\ni where $\widehat{V}^{(0)}$ is the symplectic potential, and the canonical momenta are given by Eq. \eqref{5.4}.

Using Eqs. $(\ref{eq_42})$ and $(\ref{5.1})$ we can write the symplectic potential as
\begin{align}
\widehat{V}^{(0)}&=\rho_\mu\Gamma^\mu-\frac{1}{2a^2}\vec{\pi}^2(1-\theta^{ij}\partial_jA_i)-a^2(\vec{\nabla}\cdot\vec{\Gamma}-\nabla^2A_0)^2-\vec{\pi}\cdot\vec{\nabla}\times(\vec{\nabla}\times\vec{A})+\vec{\pi}\cdot\vec{\nabla}\Gamma_0\nonumber\\
&\Big[1-\vec{\theta}\cdot(\vec{\nabla}\times\vec{A})\Big]\left[\frac{1}{2}(\vec{\Gamma}-\nabla{A_0})^2+\frac{1}{2}(\vec{\nabla}\times\vec{A})\right]\nonumber\\
&-(\vec{\theta}\cdot\vec{\Gamma}-\vec{\theta}\cdot\nabla{A_0}\Big[(\vec{\Gamma}\cdot(\vec{\nabla}\times\vec{A})-\nabla{A_0}(\vec{\nabla}\times\vec{A})^2\Big]
\nonumber\\
&+a^2\theta^{ij}\Big[\Box{A^0}-(\vec{\nabla}\cdot\vec{\Gamma})\,\partial^n\,\Big(\epsilon_{nik}(\vec{\nabla}\times\vec{A})^k(\Gamma_j+\partial_jA_0)-A_i\partial_j(\Gamma_n-\partial_nA_0)\,\Big)\Big]
\nonumber\\
&-\theta^{ij}\pi^n\pi_i\epsilon_{njk}(\vec{\nabla}\times\vec{A})^k\nonumber\\
&-\theta^{ij}\pi^n\Big[(\Gamma_i-\partial_iA_0)(\partial_n\Gamma_j-\partial_j\Gamma_n)+(\epsilon_{kil}(\vec{\nabla}\times\vec{A})^l)\partial^k((\epsilon_{njm}(\vec{\nabla}\times\vec{A})^m))\Big]\nonumber\\
&+\theta^{ij}\pi^n\Big[\Gamma_i\partial_j(\Gamma_n-\partial_nA_0)+\partial^mA_i\partial_j(\epsilon_{mnk}(\vec{\nabla}\times\vec{A})^k)\Big]+\theta^{ij}\pi^{n}A_i\partial_j\pi_n\,\,,
\label{eq_43}
\end{align}

\ni and through the Lagrangian $(\ref{eq_42})$, we have that the original symplectic variables are given by
\begin{equation}
\zeta_\alpha^{(0)}=(A_k,\rho_k,A_0,\rho_0,\Gamma_k,\pi_k,\Gamma_0),
\end{equation}

\ni where we have not written $\pi_0$ because it does not appears in the Lagrangian $(\ref{eq_42})$. So, following the FJ technique, we can identify, in Lagrangian $(\ref{eq_42})$, the non-zero canonical one-forms as 
\begin{equation}
^Aa^{(0)}_k=-\,\rho_k;\qquad^{A_0}a^{(0)}=\rho_0;\qquad^\Gamma a^{(0)}_k=-\,\pi_k\,\,.
\end{equation}

Following the symplectic formalism, we need to calculate the matrix $f_{ab}(x,y)$. Using the relations just above, we have that 

\begin{equation}
f_{ab}(x,y)=\left[\begin{array}{cc}
\mathbb{A}_{ij}&\mathbb{O}_{4\times{3}}\\
\mathbb{O}_{3\times{4}}&\mathbb{B}_{ij},
\label{eq_44}
\end{array}\right]
\end{equation}

\ni where
$$\mathbb{A}_{ij}=\left[\begin{array}{cccc}
0&\delta_{ij}&0&0\\
-\delta_{ij}&0&0&0\\
0&0&0&-1\\
0&0&1&0
\end{array}\right]
\quad
\mathbb{B}_{ij}=\left[\begin{array}{ccc}
0&\delta_{ij}&0\\
-\delta_{ij}&0&0\\
0&0&0\\
\end{array}\right].
$$

It is easy to see that the matrix $(\ref{eq_44})$ is singular. Hence, the next step is to identify the eigenvector with zero eigenvalue, which can be easily identified as being
\begin{equation}
\nu_\alpha=(\,0,0,0,0,0,0,\nu^7\,)\,\,.
\end{equation}

\ni where $\nu^7$ is arbitrary and associated with $\Gamma$.

Applying the consistency relation, in order to find new constraints, we have
\begin{equation}
{\int}dx\,dy\left((\nu^7)\,\frac{\delta\nu^{(0)}(y)}{\delta\Gamma_0(x)}\right)\,\,,
\label{eq_44.5}
\end{equation}

\ni which allows us to obtain 

\begin{equation}
{\int}dx\,\nu^7(\rho_0+\vec{\nabla}\cdot\vec{\pi})=0\,\,,
\end{equation}

\ni which is clearly a constraint, which can be denoted by 
\begin{equation}
\Omega_1(x)\equiv\rho_0+\vec{\nabla}\cdot\vec{\pi}\approx{0}\,\,,
\label{eq_45}
\end{equation}

\ni where
\begin{align}
\pi^n=&a^2\partial_{\alpha}F^{\alpha{n}}+a^2\theta^{ij}\partial^{\alpha}(F_{\alpha{i}}F^n_j-A_i\partial_jF_{\alpha}^n)+a^2\theta^{nj}(\partial_{\alpha}F^{\alpha{k}})F_{kj}+a^2\theta^{ij}\partial_j(A_i\partial_{\alpha}F^{\alpha{n}}),\nonumber\\
\rho^0=&a^2\partial_n(\partial_{\alpha}F^{\alpha{n}})+a^2\theta^{ij}\partial_n\partial^{\alpha}(F_{\alpha{i}}F^{n}_j-A_i\partial_jF_{\alpha}^n)+a^2\theta^{nj}\partial_n(\partial_{\alpha}F^{\alpha{k}}F_{kj})\nonumber\\
&+a^2\theta^{ij}\partial_{n}\Big[\partial_j(A_i\partial_{\alpha}F^{\alpha{n}})\Big]\,\,.
\end{align}

Following the FJ procedure we need to introduce this constraint into the Lagrangian $(\ref{eq_42})$ using a Lagrange multiplier $\lambda$ such that, 
\begin{equation}
\widehat{\mathcal{L}}=\pi_\mu\dot{\Gamma}^\mu+\rho_\mu\dot{A}^\mu+\dot{\lambda}\Big(\rho_0+\vec{\nabla}\cdot\vec{\pi}\Big)-\widehat{V}^{(1)}\,\,,
\label{eq_46}
\end{equation}

\ni and the new symplectic variables, observing Eq. $(\ref{eq_46})$, are
\begin{equation}
\zeta_\alpha^{(1)}=(A_k,\rho_k,A_0,\rho_0,\Gamma_k,\pi_k,\lambda).
\end{equation}

The new symplectic potential is obtained by the relation
\begin{equation}
\widehat{V}^{(1)}=\widehat{V}^{(0)}\Big|_{\Omega_1(x)=0}\,\,,
\end{equation}
so, from Eq. $(\ref{eq_43})$,
\begin{align}
\widehat{V}^{(1)}&=\rho_n\Gamma^n-\frac{1}{2a^2}\vec{\pi}^2\Big(1-\theta^{ij}\partial_jA_i\Big)-\frac{a^2}{2}\Big(\nabla\vec{\Gamma}-\nabla^2A_0\Big)^2-\vec{\pi}\cdot\vec{\nabla}\times(\vec{\nabla}\times\vec{A})\nonumber\\
&+(1-\vec{\theta}\cdot(\vec{\nabla}\times\vec{A}))\left[\frac{1}{2}(\vec{\Gamma}-\nabla{A_0})^2+\frac{1}{2}(\vec{\nabla}\times\vec{A})\right] \nonumber \\
&-(\vec{\theta}\cdot\vec{\Gamma}-\vec{\theta}\nabla{A_0})(\vec{\Gamma}\cdot(\vec{\nabla}\times\vec{A})-\nabla{A_0}(\vec{\nabla}\times\vec{A})^2)\nonumber\\
&+a^2\theta^{ij}\Big[\Box{A^0}-(\vec{\nabla}\cdot\vec{\Gamma})\,\partial^n(\epsilon_{nik}(\vec{\nabla}\times\vec{A})^k(\Gamma_j-\partial_jA_0)-A_i\partial_j(-\Gamma_n+\partial_nA_0)\Big] \nonumber \\
&-\theta^{ij}\pi^n\pi_i\epsilon_{njk}(\vec{\nabla}\times\vec{A})^k+\theta^{ij}\pi^{n}A_i\partial_j\pi_n\nonumber\\
&-\theta^{ij}\pi^n\Big[(\Gamma_i-\partial_iA_0)(\partial_n\Gamma_j-\partial_j\Gamma_n)+\Big(\epsilon_{kil}(\vec{\nabla}\times\vec{A})^l\Big)\partial^k\Big(\epsilon_{njm}(\vec{\nabla}\times\vec{A})^m \Big)\Big]\nonumber\\
&+\theta^{ij}\pi^n\Big[\Gamma_i\partial_j(\Gamma_n-\partial_nA_0)+\partial^mA_i\partial_j(\epsilon_{mnk}(\vec{\nabla}\times\vec{A})^k)\Big]
\,\,,
\end{align}

\ni and  consequently, the new vectors are
\begin{equation}
^Aa^{(1)}_k=-\rho_k;\quad^{A_0}a^{(1)}=\rho_0;\quad^{\Gamma}a^{(1)}_k=-\pi_k;\quad{\quad}^{\lambda}a^{(1)}=\rho_0+\vec{\nabla}\cdot\vec{\pi}.
\label{eq_47}
\end{equation}

The new matrix $f_{ab}(x,y)$, from Eqs. $(\ref{eq_47})$, is

$$f_{ab}(x,y)=\left[\begin{array}{cc}
\mathbb{A}_{ij}&\mathbb{D}_j\\
-\mathbb{D}_j^t&\mathbb{C}_{ij}
\end{array}\right],
$$
where
$$
\mathbb{D}_{j}=\left[\begin{array}{cccc}
0&0&0\\
0&0&0\\
0&0&0\\
0&0&1
\end{array}\right]
\quad
\mbox{and}
\quad
\mathbb{C}_{ij}=\left[\begin{array}{ccc}
0&\delta_{ij}&0\\
-\delta_{ij}&0&-\partial_j\\
0&-\partial_j&0\\
\end{array}\right].
$$

\ni and, like the matrix $(\ref{eq_44})$, it is singular.

From this last matrix, we can identify the zero modes that have non zero eigenvalues such as
\begin{equation}
\bar{\nu}_\alpha=(0,0,\bar{\nu}^3,0,\bar{\nu}_i^5,0,\bar{\nu}^7)\,\,.
\end{equation}

Consequently, using the consistency relation to find new constraint relations, we have

\begin{equation}
{\int}dx\left[\bar{\nu}^3\frac{\delta}{\delta{A_0}(x)}+\bar{\nu}_i^5\frac{\delta}{\delta{\Gamma^i}(x)}\right]{\int}dy\,\widehat{V}^{(1)}(y)\,\,,
\end{equation}

\ni which leads us to
\begin{equation}
{\int}dx\,\bar{\nu}^3(\vec{\nabla}\cdot\vec{\rho})(x)\,=\,0\,\,,
\end{equation}

\ni where we have used that $\bar{\nu}_i^5-\partial_i\bar{\nu}^3=0$. Therefore, we have that
\begin{equation}
\bar{\Omega}\equiv(\vec{\nabla}\cdot\vec{\rho})(x)=0\,\,,
\end{equation}

\ni which is a constraint.

At this point we have to introduce this constraint into the Lagrangian through the new Lagrangian multiplier, which is,
\begin{equation}
\mathcal{L}^{(2)}=\pi_\mu\dot{\Gamma}^\mu+\rho_\mu\dot{A}^\mu+\dot{\lambda}(\rho_0+\vec{\nabla}\cdot\vec{\pi})+\dot{\eta}(\vec{\nabla}\cdot\vec{\rho})-\nu^{(2)}\,\,,
\end{equation}

\ni where
\begin{equation}
\nu^{(2)}=\nu^{(1)}\big|_{\bar{\Omega}_2(x)=0}\,\,.
\end{equation}

This last relation leads us to
\begin{equation}
^Aa^{(2)}_k=-\rho_k;\quad^{A_0}a^{(2)}=\rho_0;\quad^{\Gamma}a^{(2)}_k=-\pi_k\;{\quad}^{\lambda}a^{(2)}=\rho_0+\vec{\nabla}\cdot\vec{\pi};\quad^{\eta}a^{(2)}=\vec{\nabla}\cdot\vec{\rho}\,\,,
\label{eq_48}
\end{equation}

\ni and we have new symplectic variables
\begin{equation}
\zeta_\alpha^{(0)}=(A_k,\rho_k,A_0,\rho_0,\Gamma_k,\pi_k,\lambda,\eta)\,\,.
\end{equation}

Let us write the new matrix, $f^{(2)}_{ab}(x,y)$, such as
$$f^{(2)}_{ab}(x,y)=\left[\begin{array}{cc}
\mathbb{A}_{ij}&\mathbb{E}_{j,x}\\
-\mathbb{E}_{i,y}^t&\mathbb{F}_{ij}
\end{array}\right],
$$
where
$$
\mathbb{E}_{j,x}=\left[\begin{array}{cccc}
0&0&0&0\\
0&0&0&-\partial_j\\
0&0&0&0\\
0&0&1&0\\
\end{array}\right]
\quad
\mbox{and}
\quad
\mathbb{F}_{ij}=\left[\begin{array}{cccc}
0&\delta_{ij}&0&0\\
-\delta_{ij}&0&-\partial_j&0\\
0&-\partial_j&0&0\\
0&0&0&0\\
\end{array}\right],
$$

\ni which is singular again. Now we have two zero modes, $({\bar{\nu}}_\alpha)$ and $({\nu}_\alpha)$, 
\begin{equation}
{\bar{\nu}}_\alpha=(\nu_i^{(1)},0,0,0,0,0,0,\nu_i^{(8)})\,\,,
\end{equation}
and
\begin{equation}
{\nu}_\alpha=(0,0,\nu_i^{(3)},0,\nu_i^{(5)},0,\nu_i^{(7)},0)\,\,.
\end{equation}

The next step is to apply the consistency condition in order to obtain a new constraint. However, the form of ${\nu}_\alpha$ generates the constraint $\vec{\nabla}\cdot\vec{\rho}=0$ again. Consequently, the consistency relation does not generate new constraints and the symplectic matrix remains singular.   According to the symplectic method, the model has a symmetry and the symmetry generator is the respective zero-mode. 

From the Lagrangian in Eq. (\ref{5.1}), obviously  the last term breaks the usual gauge invariance of electrodynamics  $\delta{A}_\mu=\partial_\mu\,\varepsilon$. According to the results obtained here, Podolsky electrodynamics in NC space-time is invariant by some gauge transformations, but not by the usual ones. In  \cite{afma}, we have found through the Noether method, a dual gauge invariant Lagrangian to the Lagrangian $(\ref{5.1})$.

Hence, we have to choose a convenient gauge, namely, we need to fix the gauge degrees of freedom. As in the commutative case, we will choose the generalized Coulomb gauge such as 
\begin{align}
\Big(\,1+{\Box}\,a^2\,\Big)(\vec{\nabla}\cdot\vec{A})&=0,\nonumber\\
A_0&=0,
\end{align}

\ni which leaves us with the Lagrangian
\begin{equation}
\label{AaAa}
\mathcal{L}^{(3)}=\pi_\mu\dot{\Gamma}^\mu+\rho_\mu\dot{A}^\mu+\dot{\lambda}(\rho_0+\vec{\nabla}\cdot\vec{\pi})+\dot{\eta}(\vec{\nabla}\cdot\vec{\rho})+\dot{\chi}(1-a^2\Box^2\nabla^2)\nabla\cdot\vec{A}-\widehat{V}^{(3)} \,\,,
\end{equation}

\ni where $\widehat{V}^{(3)}$ is given by
\begin{align}
\widehat{V}^{(3)}&=\rho_n\Gamma^n-\frac{1}{2}\vec{\Gamma}^2-\frac{a^2}{2}(\nabla\vec{\Gamma})^2-\frac{1}{2a^2}\vec{\pi}^2\Big(1-\theta^{ij}\partial_jA_i \Big)-\vec{\pi}\cdot \Big[\nabla\times(\nabla\times\vec{A}) \Big]\nonumber\\
&+\frac{1}{2}(\vec{\nabla}\times\vec{A})^2-\vec{\theta}\cdot(\nabla\times\vec{A}) \Big[\frac{1}{2}\vec{\Gamma}^2+\frac{1}{2}(\nabla\times\vec{A})^2 \Big]-(\vec{\theta}\cdot\vec{\Gamma}) \Big[\vec{\Gamma}\cdot(\nabla\times\vec{A}) \Big]\nonumber\\
&-a^2\theta^{ij} \Big[\partial_m\Gamma^m\partial^n(\epsilon_{nik}(\nabla\times\vec{A})^k(\Gamma_j+A_i\partial_j\Gamma_n) \Big]-\theta^{ij}\pi^n\pi_i\Big(\epsilon_{njk}(\nabla\times\vec{A})^k \Big)\nonumber\\
&-\theta^{ij}\pi^n \Big[(\Gamma_i)(\partial_n\Gamma_j-\partial_j\Gamma_n)+(\epsilon_{kil}(\vec{\nabla}\times\vec{A})^l)\partial^k((\epsilon_{njm}(\vec{\nabla}\times\vec{A})^m)) \Big]\nonumber\\
&+\theta^{ij}\pi^n\Gamma_i\partial_j\Gamma_n+\theta^{ij}\pi^n\partial^mA_i\partial_j \Big(\epsilon_{mnk}(\vec{\nabla}\times\vec{A})^k \Big)+\theta^{ij}\pi^nA_i\partial_j\pi_n.
\end{align}

The Lagrangian in \eqref{AaAa} allows us to identify the vectors 
\begin{eqnarray}
&&^Aa^{(3)}_k=-\rho_k\qquad;{\quad}^{A_0}a^{(3)}=\rho_0\qquad;{\quad}^{\Gamma}a^{(3)}_k=-\pi_k\;; \nonumber \\
\mbox{} \nonumber \\
&&{\,}^{\lambda}a^{(3)}=\rho_0+\vec{\nabla}\cdot\vec{\pi}\quad;{\quad}^{\eta}a^{(3)}=\vec{\nabla}\cdot\vec{\rho}\quad;{\quad}^{\chi}a^{(3)}=\nabla^2_p\,({\nabla}\cdot\vec{A})\;\;,
\end{eqnarray}

\ni and concerning the third symplectic matrix we have that
$$
f^{(3)}_{ij}=\left[\begin{array}{cccccccc}
0&\delta_{ij}&0&0&0&0&0&-\nabla^2_p\partial^x_j\\
-\delta_{ij}&0&0&0&0&0&-\partial^x_j&0\\
0&0&0&0&0&1&0&0\\
0&0&0&0&\delta_{ij}&0&0&0\\
0&0&0&-\delta_{ij}&0&\partial^x_i&0&0\\
0&0&1&0&\partial^x_i&0&0&0\\
0&\partial^y_i&0&0&0&0&0&0\\
\nabla^2_p\partial^x_j&0&0&0&0&0&0&0\\
\end{array}\right].
$$

\ni which is non singular.   We have determined, after a long algebraic work, its inverse and consequently we obtained the Dirac brackets of the theory. Thus, the Dirac brackets are
\bee
\label{dirac1}
\Big\{A_k(x),\rho^m(y)\Big\}_{DB}=\delta^m_k\delta(x,y)-\nabla^2_p\partial_k\partial^mG(x,y) 
\eee

\ni and  
\bee
\label{dirac2}
\Big\{\Gamma_k(x),\pi^m(y)\Big\}_{DB}=\delta^m_k\delta(x,y)\,\,, 
\eee

\ni where $G(x,y)$ is the Green function
\begin{equation}
\nabla^2_p\,\Big(\nabla^2G(x,y)\,\Big)=\delta^{(3)}(x,y)\,\,.
\end{equation}

\ni These generalized brackets can be used to map the classical theory into a quantum system.

\section{Conclusion}

The study of theories in NC space-time has motivated an enormous number of papers through the last years and the min target is the attempt to understand the Planck scale physics such as the early Universe physics.  In other words, to introduce NC characteristics into a classical or a quantum physics is to introduce a Planck scale parameter having an area dimension which can be constant or not.  In our case in this work, the NC parameter is constant which constitutes the so-called canonical noncommutativity.

Although, as we have said, NC space-time theories are been intensively explored, it is natural to exist some gaps that are not properly or totally investigated in any kind of issue in theoretical physics.  Some aspects such as the constraints aspects such as the Dirac-Bergmann approach to analyze constrained systems is one of these aspects that were not explored sufficiently until now.  And this is exactly the issue that we have investigated in this work since the gauge invariance subject is directly connected to this constrained analysis.
An alternative to Dirac-Bergmann approach is the well known Faddeev-Jackiw technique which, in an intermediate step, obtain the Dirac brackets of the system.   

Another interesting aspect that were not sufficiently explored so far is the higher-derivatives theories in NC space-times. 
In this work we have analyzed both of these aspects, namely, the NC version of the Podolsky electromagnetism is an example of a constrained system that has higher-derivatives.  Another aspect is that we have used the FJ approach to attack this problem.

The main issue here was to use the FJ method in order to  obtain the Dirac brackets of the Podolsky electrodynamics described in NC space-time. As a constrained system, the Dirac brackets of NC Podolsky electrodynamics cannot be obtained via canonical way because of the inconsistencies that appear when we try to promote the classical quantities to operators. The FJ method has showed itself to be an economical way to deal with a higher-order derivatives Lagrangian. 

To start, we have reviewed the main aspects of both the FJ formalism and  the Podolsky electrodynamics in NC space-time. After that, we have applied the FJ method in this NC Podolsky electrodynamics and after some algebraic work we obtained the Dirac brackets of this model, which is the main result of this paper.

We believe that the relevance of these results is based on the fact that we have now a new point of view concerning constrained systems depicted in NC space-times and, consequently, the FJ method, in the light of NC space-time theories. In many cases, the FJ procedure is an economical alternative to work with constrained systems as we have shown in this paper.    



\section{Acknowledgments}

\ni E.M.C.A.  thanks CNPq (Conselho Nacional de Desenvolvimento Cient\' ifico e Tecnol\'ogico), Brazilian scientific support federal agency, for partial financial support, Grants numbers 302155/2015-5 and 442369/2014-0 and the hospitality of Theoretical Physics Department at Federal University of Rio de Janeiro (UFRJ), where part of this work was carried out.  



\end{document}